\documentclass[twocolumn,showpacs,preprintnumbers,amsmath,amssymb]{revtex4}
\usepackage[dvips]{graphicx}
\usepackage{dcolumn}

\begin{document}

\title
{Impurity-induced broadening of the transition to a Fulde-Ferrell-Larkin-Ovchinnikov phase
}

\author{Ryusuke Ikeda}

\affiliation{%
Department of Physics, Kyoto University, Kyoto 606-8502, Japan
}

\date{\today}


\begin{abstract} 
Recent study on doping effects in the heavy fermion superconductor CeCoIn$_5$ has shown that a small amount of doping induces unexpectedly large broadening of the transition into the high field and low temperature (HFLT) phase of this material. To resolve this observation, effects of quenched disorder on the second order transition into a longitudinal Fulde-Ferrell-Larkin-Ovchinnikov (FFLO) state are examined. The large broadening of the transition is naturally explained as a consequence of softness of each FFLO nodal plane. The present results strongly support the scenario identifying the HFLT phase of CeCoIn$_5$ with a longitudinal FFLO state. 
\end{abstract}



\maketitle


Understanding the high field and low temperature (HFLT) phase \cite{Bianchi,Watanabe,Kenzel} realized in the heavy-fermion superconductor CeCoIn$_5$ in the parallel field configuration is an issue under hot debate on unconventional superconductivity. Based on a large paramagnetic depairing of this material and on a close examination on the phase diagram, the HFLT phase has been identified \cite{Bianchi,Watanabe} with a Fulde-Ferrell-Larkin-Ovchinnikov (FFLO) \cite{FF,LO} vortex lattice with a modulation parallel to the field \cite{RI072}, which will be dubbed hereafter the longitudinal FFLO state. This FFLO scenario naturally explains most of properties relevant to the HFLT phase, such as the striking reduction of the vortex tilt modulus in the HFLT phase \cite{Watanabe,RI071} and the second order character of the transition (HAT) between the HFLT and the ordinary Abrikosov lattice \cite{Abrikosov} phases. On the other hand, following a recent observation of a transverse \cite{Kenzel} antiferromagnetic (AFM) order, much attention have been paid to magnetic properties and have led to an alternative scenario against the FFLO one, in which HAT is identified with the onset of a $Q$-phase, which is a pair density wave with {\it no longitudinal} modulation and accompanied by a {\it bulk} AFM order \cite{Agter,Little} (See also Ref.11). It has been speculated \cite{RI09} and demonstrated \cite{Yanase}, however, that the observed AFM order may be an event localized around the transverse nodal planes in the longitudinal FFLO state \cite{RI09}. Besides this AFM ordering, a striking doping effect on HAT has been reported in heat capacity measurements \cite{Tokiwa,Roman}: For both of AFM dopants (Hg and Cd) and a nonmagnetic (Sn) one, even an extremely small amount of doping has induced a transition broadening and a dramatic reduction of the heat capacity jump at HAT, suggesting that an ordering occurring through HAT is highly fragile. This is qualitatively and quantitatively different from the familiar {\it electronic} impurity effects on a unconventional pairing transition \cite{AG} and a FFLO transition \cite{VVG} in which just a shift of the transition point is expected. 

In this work, we examine quenched disorder effects on HAT between the longitudinal FFLO and the ordinary Abrikosov vortex lattices and explain the observed broadening of the heat capacity near and below HAT \cite{Tokiwa,Roman} with the help of numerical details of a microscopically derived Ginzburg-Landau (GL) Hamiltonian \cite{RI071}. A soft tilt rigidity \cite{RI071} of the FFLO nodal plane is found to be the main origin of the dramatic broadening of heat capacity curves \cite{Tokiwa,Roman}. For comparison, the same analysis is also performed for a GL model appropriate for the $Q$-phase scenario \cite{Agter,Little,Mitrovic} of the HFLT phase. In the latter, the primary effect of quenched disorder is always a simple shift of the transition point unaccompanied by a notable impurity-induced broadening. Based on these results, the validity on the picture \cite{Bianchi,Watanabe,RI072} identifying the HFLT phase with the longitudinal FFLO state is stresseed. 

We consider the Hamiltonian ${\cal H} = N(0) (2 \pi \xi_0)^3 T_c^2 \, [ h_0 + h_p]$, where 
\begin{eqnarray}
h_0 &=& \int d^3r \biggl[ \alpha Q^2({\bf r}) + \frac{\beta}{2} Q^4({\bf r}) + \varepsilon \sum_{i \neq z} (\partial_i s({\bf r}))^2 \nonumber \\ 
&+& \frac{1}{2} ( \, u_i \Pi_{ij} u_j + \gamma \, Q^2 {\rm div}{\bf u} \, ) \biggr] \label{GL}
\end{eqnarray} 
and 
\begin{equation}
h_p = \int d^3r \biggl[ 2 h({\bf r}) Q_0 s({\bf r}) + {\bf f}\cdot{\bf u} \biggr]
\end{equation}
are dimensionless and valid in the Larkin-Ovchinnikov (LO) {\it vortex} state with the pair-field (superconducting order parameter) $\Delta({\bf r}) = \sqrt{2} \Delta(x, y) {\rm cos}(Q_0 z + s({\bf r}))$ near HAT. Here, $N(0)$ is the density of states on the Fermi surface in the normal state, $T_c$ is the zero field superconducting transition temperature of the undoped system, $Q=Q_0 + \partial_z s$ is the FFLO order parameter expressing the inverse of the local period of FFLO modulation parallel to the applied field ${\bf H} \parallel {\hat z}$, ${\bf s}= s {\hat z}$ is the displacement field of the nodal planes lying in the $x$-$y$ plane in equilibrium, ${\bf u} \perp {\hat z}$ is the compressional displacement of the vortex lattice arising from $\Delta(x, y)$, $\Pi_{ij}$ is an elastic matrix of vortices to be defined later, and $\beta$, $\varepsilon > 0$. In eq.(\ref{GL}), the first two terms describe the mean field ordering of the longitudinal FFLO state, while other terms expressing the elasticity of the nodal planes and the vortices and the coupling between them have been examined elsewhere \cite{RI071}. Any length was already normalized by $2 \pi \xi_0$, where $\xi_0$ is the coherence length in $T=0$ limit. In calculating the heat capacity, numerical data \cite{RI071} of the dimensionless coefficients $\alpha$, $\gamma$, and $\varepsilon$ will be used (see below) \cite{com}. Among possible roles of impurities, we focus hereafter on their quenched disorder effects on the order parameter fields described by eq.(2), and the electronic impurity effects will be commented on at the end of this paper. The random field terms in eq.(2), for instance, follow from the conventional random $T_c$ term $\sim \int_{\bf r} w({\bf r}) |\Delta({\bf r})|^2$ in the superconducting GL Hamiltonian \cite{RI072}, and the presence of the factor $Q_0$ in the nodal plane pinning term proportional to $h({\bf r})$ may be justified from the model $w({\bf r}) = \sum_\nu w_\nu \delta^{(3)}({\bf r} - {\bf r}_\nu)$. A possible randomness of $\alpha$ implying spatial inhomogenuities of the HAT temperature is of a higher order compared with the $h$-term and thus, was neglected. 

To examine the free energy density ${\cal F}$ in the FFLO state near HAT, thermal fluctuations of $s$ and ${\bf u}$ will be neglected. Then, the method used by Larkin and Ovchinnikov \cite{LO2} for a second order transition will be adopted and extended here in a self consistent manner to obtain ${\cal F}$. After taking variations of ${\cal H}$ with respect to $Q_0$, ${\bf u}$, and $s$ and keeping the contributions up to O($s^2$) in the $Q^4$ term, the variational equations 
\begin{eqnarray}
\alpha Q &+& \beta Q_0^3 + \biggl( h + \frac{\gamma}{2} (\Pi^{-1})_{ij} \partial_z \partial_i f_j \biggr) s  = - Q_0 \biggl[ 3 \beta (\partial_z s)^2 \nonumber \\
&-& \frac{\gamma^2}{2} \sum_{i,j \neq z} (\partial_i \partial_z s) (\Pi^{-1})_{ij} (\partial_j \partial_z s) \biggr], 
\nonumber \\
\label{var1}
\end{eqnarray}
and 
\begin{eqnarray}
[(\alpha &+& \, 3 \beta Q_0^2 \,) \partial_z^2 + Q_0^2 \gamma^2 (\Pi^{-1})_{ij} \partial_z^2 \partial_i \partial_j + \varepsilon \, \partial_\perp^2] s \nonumber \\ 
&=& Q_0 \biggl( h + \frac{\gamma}{2} (\Pi^{-1})_{ij} \partial_z \partial_i f_j \biggr)
\label{var2}
\end{eqnarray}
follow. By taking the random average of eq.(\ref{var1}), the self consistent equation determining $Q_0$ 
\begin{equation}
\beta Q_0^2 = \frac{ I_1 - \alpha}{ 1 + 3 I_2 } 
\label{self}
\end{equation}
is obtained, where 
\begin{eqnarray}
I_1 &=& \int\frac{d^3k}{(2 \pi)^3} \frac{\delta_0}{D_k} , \nonumber \\
I_2 &=& \int\frac{d^3k}{(2 \pi)^3} \delta_0 \biggl( 1 - \frac{\gamma^2}{6 \beta} \Pi^{-1}(k) k_\perp^2 \biggr) D_k^{-2}, 
\label{integ}
\end{eqnarray}
and 
$
D_k = [I_1 + \beta (2 - 3 I_2- \gamma^2 \Pi^{-1}(k) k_\perp^2/\beta) Q_0^2 ] k_z^2 + \varepsilon \, k_\perp^2$. 
Here, $\Pi_{ij}({\bf k})$ was isotropized in the way $\delta_{ij} \Pi$ with $\Pi=H^2 k^2/(4 \pi N(0) T_c^2)$, $\delta_0$ denotes the random average of $h_{\bf k} h_{-{\bf k}} + \Pi^{-2} k_z^2 |{\bf k}\cdot{\bf f}_k|^2/8$, and, for brevity, its $|{\bf k}|$ dependence has been neglected. After the ${\bf k}$-integrals in eq.(\ref{integ}), an upper momentum cutoff $k_c$ of order unity will be absorbed into the bare disorder strength via its redefinition, $\delta_0 k_c/(4 \pi^2) \to \delta_0$. 
The resulting expressions can be simplified further because the $\gamma$-dependent terms appearing through the ${\bf u}$-variation are quantitatively negligible. In fact, using typical values of $\gamma$ ($\sim 0.015$) and $\beta$ ($\sim 0.004$) obtained elsewhere \cite{RI071,RI072} through the study of the phase diagram of CeCoIn$_5$ near $H_{c2}(0)$, $\gamma^2 N(0) 4 \pi T_c^2/(6 H^2 \beta)$ is estimated at most as $\sim 10^{-2} k_{\rm B} T_c/E_{\rm F}$, where $E_{\rm F}$ is the Fermi energy. For this reason, any $\gamma$-dependent terms will be neglected below. Then, we have 
\begin{equation}
I_1 = \frac{2 \delta_0}{\varepsilon} \biggl(\frac{m}{\varepsilon} - 1 \biggr)^{-1/2} {\rm tan}^{-1}\biggl(\sqrt{\frac{m}{\varepsilon} - 1} \biggr),
\label{I1}
\end{equation}
and $I_2 = -\partial I_1/\partial m$, where $m=I_1 + (2 - 3I_2) \beta Q_0^2$. 

It should be stressed that, although effects of the vortex displacement are negligible, the presence of the field-induced vortices is not negligible because the nonvanishing $\varepsilon$ ($\sim 0.002$) \cite{RI071} used in our analysis is a consequence of the orbital pair-breaking. On the other hand, the soft nodal plane implied by such a small $\varepsilon$ is a reflection of the large paramagnetic depairing in CeCoIn$_5$ and, in the impure case, has a crucial impact on the thermodynamics near HAT. Note also that the nodal plane is softer in more anisotropic systems such as the organic superconductors because a mass anisotropy enhances the paramagnetic effect in the same field configuration. 

Based on the expressions obtained above, the change of free energy accompanying the FFLO ordering follows simply from the random-average of ${\cal H}$, and ${\cal F}$ is given by 
\begin{equation}
{\cal F} = - \frac{1}{2 \beta} N(0) T_c^2 \frac{ 1 + 6 I_2}{(1 + 3 I_2 )^2} (\alpha - I_1)^2.
\label{free}
\end{equation}
The resulting heat capacity $C(T)$ curves below $T_s(\delta_0)$ are given in Fig.1 as a function of $\delta_0$, where the heat capacity is normalized by its value at $T_s$ in the pure ($\delta_0=0$) case in the form ${\overline C}(T) \equiv C(T)/C(T=T_s, \delta_0=0)$, and $T_s(\delta_0)$ is the HAT temperature. The fact that ${\overline C}(T)$ is significantly broadened and depressed by a small amount of disorder seems to be consistent with the feature seen around HAT in CeCoIn$_5$ \cite{Tokiwa,Roman}. According to eq.(\ref{I1}), this sensitivity of $C$ to quite a small $\delta_0$ is a consequence of the fact that the {\it effective} disorder strength is not $\delta_0$ but $\delta_0/\varepsilon$. A smaller $\varepsilon$ enhances the disorder effect and leads to a more dramatic broadening of the transition. Further, note that a large depression of ${\overline C}$-value also implies that the period $\sim Q_0^{-1}$ of the FFLO modulation remains macroscopically long even at lower temperatures. 

Two crucial features are seen in this figure in relation to the broadening : First, $T_s$ increases with $\delta_0$ up to $\sim 10^2 \delta_0 T_c/\varepsilon$, and further, a broad peak appears in ${\overline C}(T)$ which, as far as $\delta_0/\varepsilon > 2.5 \times 10^{-5}$, lies {\it much below} $T_s$. In fact, this broad peak occurs in the region where $m > \varepsilon$, and hence, tranverse spatial variations ($|{\bf k}_\perp| \gg |k_z|$) of the nodal planes are dominant. Thus, the broadening accompanied by the suppression of the peak in Fig.1 is a consequence of the softness of each nodal plane. We note that the present approach is not applicable to the so-called critical region in the close vicinity of $T_s$ because of the neglect of nonlinear corrections in eq.(2), which may play important roles when $m \ll \delta_0/\varepsilon$, or equivalently $|T_s - T| \ll 1.5 \times 10^{-2} \, T_c$. In fact, other nonperturbative approaches are necessary to describe the behaviors in $T > T_s$ including the Griffiths regime \cite{Dotsenko}. However, the disorder-induced broadening, which is our main focus, is a feature far below the critical region and thus, is believed to be reasonably captured within the present approach. 

\begin{figure}[t]
\scalebox{0.3}[0.3]{\includegraphics{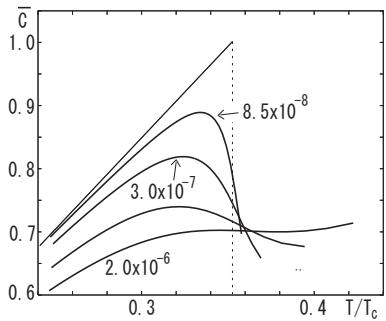}}
\caption{ Results of normalized heat capacity ${\overline C}(T)$ (thick solid curves) below $T_s(\delta_0)$ and at a fixed magnetic field following from eq.(\ref{free}) for  $\delta_0=8.5 \times 10^{-8}$ (top), $3.0 \times 10^{-7}$, $1.0 \times 10^{-6}$, and $2.0 \times 10^{-6}$ (bottom). Data in Ref.\cite{RI071} on the coefficients in eq.(1) at $H=0.5 H_{\rm orb}^{({\rm 2D})}(0)$ have been used, where $H_{\rm orb}^{({\rm 2D})}(0) = 0.56 \phi_0/(2 \pi \xi_0^2)$ is the 2D orbital-limiting field, and $\phi_0$ is the flux quantum. The right end of each curve corresponds to the result at each $T_s(\delta_0)$. The thin solid line denotes ${\overline C}(T)$ below $T_s = 0.354 T_c$ in the pure ($\delta_0=0$) case.}
\label{fig.1}
\end{figure}

\begin{figure}[b]
\scalebox{0.3}[0.3]{\includegraphics{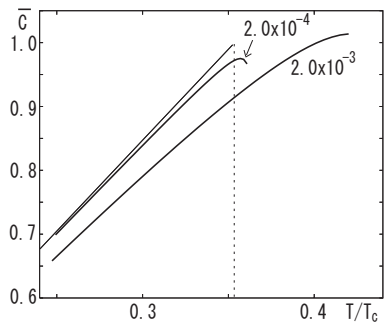}}
\caption{Results corresponding to Fig.1 in the $Q$-phase scenario. The same $\alpha$ as in Fig.1 is used, while $\gamma_+$ is $T$-independent. Although $T_s$ of the $\delta'_0=2.0 \times 10^{-3}$ curve is almost the same as that of the $\delta_0 = 2.0 \times 10^{-6}$ one in Fig.1, no notable broadening is seen in this figure.}  
\label{fig.2}
\end{figure}

To explain whether or not the impurity-induced broadening seen in Fig.1 is peculiar to the FFLO to Abrikosov lattice transition, it is instructive to repeat the same analysis in the $Q$-phase scenario \cite{Agter,Little,Mitrovic} 
on HAT of 
CeCoIn$_5$ in which the HFLT phase is identified with a spin-triplet pair-density wave accompanied by an AFM order with a {\it transverse} modulation wavevector ${\bf Q}$. In this case, after being minimized with respect to the coupled AFM order, the resulting GL Hamiltonian takes a form typical of a "two band" superconductor expressed by the $d$-wave pair field and a triplet pair field $\Psi$ with {\it no} ${\bf Q}$-modulation, in which the ordered $d$-wave vortex solid plays roles of a periodic potential for $\Psi$. Thus, an energy cost due to a mismatch between the vortex positions of the $d$-wave pair field and $\Psi$ leads to a gradient term for the latter. If we only have to focus on the amplitude fluctuation $\psi_+$ of $\Psi$ (see below), the resulting GL Hamiltonian for slowly varying components of $\Psi$ is equivalent to the familiar random $T_c$ model for the Ising spin system and is expressed by 
\begin{eqnarray} 
{\cal H}_u &=& N(0) (2 \pi \xi_0)^3 T_c^2 \int d^3r \biggl[ \alpha \Psi_0^2 + \frac{\beta'}{2} \Psi_0^4 + (\alpha \nonumber \\ 
&+& 3 \beta' \Psi_0^2) \psi_+^2  
+ \gamma_+ (\nabla \psi_+)^2 + 2 h' \Psi_0  \psi_+  \biggr] 
\label{agt}
\end{eqnarray}
to the leading orders in $\psi_+$, where $\Psi_0$ is a real and constant amplitude of the equilibrium lattice solution of $\Psi$, and the length scales were isotropized. The phase fluctuation of $\Psi$ neglected above is that of "interband" Josephson phase which remains massive \cite{Tony}. It leads only to a small $T$-independent contribution to $\alpha$ in the first term of eq.(\ref{agt}), which can safely be omitted in examining the heat capacity. Equation (\ref{agt}) may be obtained in the Abrikosov's mean field analysis \cite{Abrikosov} and its extension \cite{Eilenberger} focusing on the lowest Landau level modes of $\Psi$. Then, the expression corresponding to eq.(\ref{self}) can easily be obtained, which becomes 
\begin{equation}
\beta' \Psi_0^2 = \frac{ I'_1 - \alpha}{1 + 3 I'_2}, 
\end{equation}
where the integrals $I'_n$ ($n=1$, $2$) are given by $I'_1 = \delta'_0 (1 - \pi (m'/(\gamma_+ k_c^2))^{1/2}/2)$ with an upper momentum cutoff $k_c$, and $I'_2 = - \partial I'_1/\partial m'$, where $\delta'_0$ is the random average of $h'_{\bf k} h'_{-{\bf k}}$ multiplied by $k_c^3/(2 \pi^2)$, and $m'= \alpha + 3 \beta' \Psi_0^2$. In addition, the free energy density is given by eq.(\ref{free}) with $I_n$ replaced by $I'_n$. The resulting heat capacity curves in the $\Psi_0 \neq 0$ state are given in Fig.2. Although, for simplicity, the combination $\gamma_+ k_c^2$ has been set to be unity in Fig.2, qualitative results are independent of the $\gamma_+$-value because it can be absorbed into the effective value of disorder strength. The figure shows that no dramatic broadening occurs even for a moderate strength of disorder, and a deviation from the pure ($\delta'_0=0$) case is noticable just near the critical region. Clearly, the main effect of disorder is to shift the transition point.  

Figures 1 and 2 show that the transition broadening induced by a small amount of dopings in CeCoIn$_5$ \cite{Tokiwa,Roman} is consistent only with the case in which its HFLT phase is a vortex lattice with a longitudinal modulation. As already mentioned, the broadening is seen experimentally irrespective of the type of the doped element. This fact indicates that the broadening should be ascribed not to a change of electronic details but rather to a quenched disorder effect due to the dopants. 

On the other hand, the measurements show that dependences of the nominal position of HAT on the type of dopants are nonuniversal : For the magnetic doping such as Hg and Cd, the position of the broad peak of $C(T)$ is shifted rather to higher temperatures with doping \cite{Tokiwa}, while the doping of the nonmagnetic element Sn has shifted the broad peak to lower temperatures \cite{Roman}. According to Fig.1, however, the peak position lies markedly below the actual transition point $T_s(\delta_0)$, and this deviation between those two temperatures is enhanced with increasing $\delta_0$. Besides this, the conventional electronic impurity effect due to a nonmagnetic doping induces a suppression of the transition temperature into an unconventional pairing state \cite{AG} or the FFLO state \cite{AI}. Incorporating this electronic effect on $T_s$ in eq.(1) is easily performed simply by redefining $\alpha$ and does not affect the transition broadening in Fig.1 induced by quenched disorder. Then, due to the coexistence of such two impurity effects on $T_s$ competitive with each other, it is not easy to predict the actual HAT point in real materials. At least, the reduction of the broad peak position due to the Sn doping \cite{Roman} suggests the presence of a remarkable $T_s$-reduction of an electronic origin \cite{AI}. For the magnetic doping \cite{Tokiwa}, however, the broad peak position seems to increase with doping, suggestive of the presence of an electronic mechanism leading to a slight increase of $T_s$. A separate study on {\it electronic} details is needed to explain the increase of the broad peak position and of the corresponding $H_{c2}(T)$ \cite{Roman} due to a magnetic doping. 

The present result also has implication on the issue of FFLO phases in strongly anisotropic organic superconductors \cite{organic1}. In these materials, a longitudinal FFLO state like that in CeCoIn$_5$ has not been observed, and instead, only a transverse modulation in the plane perpendicular to the field has been argued to appear \cite{organic1}. Theoretically, however, the possible FFLO {\it vortex} state at higher temperatures is predicted to be the longitudinal one \cite{RI072}. The present result leads to the conjecture that, due to the strong anisotropy, the $\varepsilon$-value in those systems which are closer to the vortex-free Pauli limit than CeCoIn$_5$ is extremely low so that the transition into the longitudinal FFLO state is dramatically broadened and becomes invisible due only to a small amount of impurities. In other words, the period of the spatial modulation parallel to the field remains significantly long even at lower temperatures. Further discussion on organic materials based on the present picture will be given elsewhere. 

In conclusion, the doping-induced large broadening of the transition between HFLT and the ordinary Abrikosov lattice phases in CeCoIn$_5$ definitely shows that, in contrast to the proposed $Q$-phase, the HFLT phase has a spatial modulation parallel to the field, and that the longitudinal FFLO vortex lattice is its best candidate of such a highly fragile HFLT phase. The present result is also relevant to the issue on FFLO phases to be observed in organic materials in high fields. 

This work was supported by Grant-in-Aid for Scientific Research [Grants No. 20102008 and No. 21540360] from MEXT, Japan.


\begin{thebibliography}{99}

\bibitem{Bianchi} A. D. Bianchi et al., Phys. Rev. Lett.{\bf 91} 187004 
(2003). 
\bibitem{Watanabe} T. Watanabe et al., Phys. Rev. B {\bf 70}, 020506(R) 
(2004). 
\bibitem{Kenzel} M. Kenzelman et al., Science {\bf 321}, 1652 (2008). 
\bibitem{FF} P. Fulde and R. A. Ferrell, Phys. Rev. {\bf 135}, A550 (1964). 
\bibitem{LO} A. I. Larkin and Yu. N. Ovchinnikov, Sov. Phys. JETP {\bf 20}, 762 (1965). 
\bibitem{RI072} R. Ikeda, Phys. Rev. B {\bf 76}, 134504 (2007). 
\bibitem{RI071} R. Ikeda, Phys. Rev. B {\bf 76}, 054517 (2007). 
\bibitem{Abrikosov} A. A. Abrikosov, Sov. Phys. JETP {\bf 5}, 1174 (1957). 
\bibitem{Agter} D.F. Agterberg, M. Sigrist, and H. Tsunetsugu, Phys. Rev. Lett. {\bf 102}, 207004 (2009). 
\bibitem{Little} A. Aperis et al., J. Phys. Condens.Matter {\bf 20}, 434235 (2008). 
\bibitem{Mitrovic} G. Koutroulakis et al., arXiv:0912.3548. 
\bibitem{RI09} R. Ikeda, Phys. Rev. Lett. {\bf 102}, 069703 (2009). 
\bibitem{Yanase} Y. Yanase and M. Sigrist, J. Phys. Soc. Jpn. {\bf 78}, 114715 (2009). 
\bibitem{Tokiwa} Y. Tokiwa et al., Phys. Rev. Lett. {\bf 101}, 037001 (2008). 
\bibitem{Roman} R. Movshovich et al., presented in M$^2$S 2009. 
\bibitem{AG} A. A. Abrikosov and L. P. Gor'kov, Sov. Phys. JETP {\bf 12}, 1243 (1961). 
\bibitem{VVG} A. B. Vorontsov, I. Vekhter, and M. J. Graf, Phys. Rev. B {\bf 78}, 180505 (2008). 
\bibitem{com} In the notation in Ref.\cite{RI071}, $\alpha = c^{(2)} (T_c \gamma^{1/2} r_B/v_F)^2$, and $\varepsilon=E_2/2$. 
\bibitem{LO2} A. I. Larkin and Yu. N. Ovchinnikov, Sov. Phys. JETP {\bf 34}, 651 (1972). 
\bibitem{Dotsenko} G. Tarjus and V. Dotsenko, J. Phys. A {\bf 35}, 1627 
(2002).  
\bibitem{Eilenberger} G. Eilenberger, Phys. Rev. {\bf 164}, 628 (1967). 
\bibitem{Tony} A. J. Leggett, Prog. Theor. Phys. {\bf 36}, 901 (1966). 
\bibitem{AI} H. Adachi and R. Ikeda, Phys. Rev. B {\bf 68}, 184510 (2003). 
\bibitem{organic1} S. Uji et al., Phys. Rev. Lett. {\bf 97}, 157001 (2006). 
\end{thebibliography}
\end{document}